\documentclass[11pt,a4paper]{article}
\pdfoutput=1
\usepackage{jheppub}
\usepackage{caption}
\usepackage{subcaption}
\usepackage{comment}
\usepackage{color}
\newcommand{\fig}[1]{Fig.\ \ref{#1}}
\newcommand{\figtwo}[2]{Fig.\ \ref{#1} (\subref{#2})}
\newcommand{\refeq}[1]{Eq.~(\ref{#1})}

\newcommand{\sect}[1]{Section #1}

\newcommand{\be}{\begin{equation}}
\newcommand{\ee}{\end{equation}}
\newcommand{\bea}{\begin{eqnarray}}
\newcommand{\eea}{\end{eqnarray}}
\newcommand{\nn}{\nonumber}

 \def\cB{{\cal B}} 
  \def\cF{{\cal F}}
 \def\cH{{\cal H}}

\newcommand{\eq}[1]{(\ref{#1})}

\begin{document}
\begin{titlepage}
\thispagestyle{empty}

\vspace{2cm}
\begin{center}
\font\titlerm=cmr10 scaled\magstep4
\font\titlei=cmmi10
scaled\magstep4 \font\titleis=cmmi7 scaled\magstep4 {\Large{\textbf{ Jets in a strongly coupled anisotropic plasma}
\\}}
\setcounter{footnote}{0}
\vspace{1.5cm} \noindent{{
 Kazem Bitaghsir Fadafan$^{a,b}$\footnote{e-mail:bitaghsir@shahroodut.ac.ir, K.Bitaghsir-Fadafan@soton.ac.uk}, Razieh Morad$^{c}$\footnote{e-mail:razieh.morad@uct.ac.za}
}}\\
\vspace{0.8cm}

{\it $^{a}$ Faculty of Physics, Shahrood University of Technology, P.O.Box 3619995161 Shahrood, Iran\\}
{\it $^{b}$ STAG Research Centre  Physics and Astronomy,
 University of Southampton, Southampton, SO17 1BJ, UK\\}
{\it $^{c}$ Department of Physics, University of Cape Town, Private Bag X3, Rondebosch 7701, South Africa}

\vspace*{.4cm}
\end{center}
\vskip 2em
\setcounter{footnote}{0}
\begin{abstract}

In this paper, we study the dynamics of the light quark jet moving through the static, strongly coupled $\mathcal{N}=4$,  anisotropic plasma with and without charge. The light quark is presented by a 2-parameters point-like initial condition falling string in the context of the AdS/CFT.  We calculate the stopping distance of the light quark in the anisotropic medium and compare it with its isotropic value. We study the dependency of the stopping distance to the both string initial conditions and background parameters such as anisotropy parameter or chemical potential. Although the typical behavior of the string in the anisotropic medium is similar to the one in the isotropic AdS-Sch background, the string falls faster to the horizon depending on the direction of moving. Particularly, the enhancement of quenching is larger in the beam direction. We find that the suppression of stopping distance is more prominent when the anisotropic plasma have the same temperature as the isotropic plasma. 

\end{abstract}
\end{titlepage}
\tableofcontents

\section{Introduction}

One of the most striking features of experiments at the Relativistic Heavy Ion Collider (RHIC) and the Large Hadron Collider (LHC) is the collective flow and its hydrodynamic interpretation \cite{Ackermann:2000tr,Adcox:2002ms,Adler:2003kt} which has been studied both experimentally and theoretically \cite{Kolb:2003dz}. These studies show that the produced medium in the heavy ion collisions equilibrates very fast and forms some kind of fluid. Also because of the collective flow, the medium would be anisotropic and not static \cite{ATLAS:2012at, Armesto:2004pt}. High energy partons which are created at the early stages of the collision, traverse the hot and dense matter, radiate gluons and lose their energies into the medium. This phenomena which is known as jet quenching \cite{Mehtar-Tani:2013pia}, is a good tool to study such medium \cite{Gyulassy:1993hr,Baier:1996sk,Zakharov:1997uu,Wiedemann:2000za}.

There are lots of evidences indicate the quark-gluon plasm (QGP) behaves as a strongly coupled fluid \cite{Shuryak:2004cy}. The fact that the lattice QCD is not an appropriate tool for understanding the time evolution of such strongly coupled plasma provides a strong motivation to use the gauge-gravity duality \cite{CasalderreySolana:2011us}. This duality equates certain gauge theories with quantum gravity theories in one extra dimension \cite{Maldacena:1997re,Witten:1998qj,Gubser:1998bc}. The extra dimension is interpreted as the renormalization group scale.

Although the ideal hydrodynamics successfully describes many properties of QGP, further studies propose that the medium must be anisotropic in the early times during the formation of QGP \cite{Florkowski:2008ag, Florkowski:2009sw, Ryblewski:2008fx, Romatschke:2003ms, Arnold:2005vb, Rebhan:2008uj}. These models proposed that within this time period, the pressure of the medium along the transverse direction may surpasses the pressure along the longitudinal direction. Different aspects of anisotropic QGP using the hard-loop effective field theory \cite{Mrowczynski:2004kv} have been studied, such as momentum broadening of jets \cite{Romatschke:2006bb,Baier:2008js}.

The anisotropic geometries have been studied using the gauge/gravity duality in \cite{Mateos:2011ix, Janik:2008tc, Giataganas:2017koz}. The bottom-up anisotropic background proposed by Janik and Witaszczyk \cite{Janik:2008tc} results in an oblate or prolate geometry depends on the parameters of the model. This model has a naked singularity in the bulk geometry. On the other hand, the gravity dual to the top-down model of Mateos and Trancanelli (MT) \cite{Mateos:2011ix, Mateos:2011tv} is regular and the dual theory is an anisotropic $\mathcal{N} = 4$ SYM plasma with rotational symmetry in the $x- y$ directions, while the $z$ direction corresponds to the beam direction. The anisotropy is introduced by deforming the gauge theory with a parameter, $\theta$ which depends on the $z$ coordinate.

Using these holographic models, drag force, jet quenching parameter and quarkonium dissociation in the anisotropic medium were calculated \cite{Chernicoff:2012iq,Giataganas:2012zy,Chernicoff:2012gu,Bellantuono:2017msk}. The jet quenching parameter, $\hat{q}$ the momentum broadening of a highly relativistic parton moving through a non-Abelian plasma, in the anisotropic background has been studied in \cite{Chernicoff:2012gu}. It is shown that the $\hat{q}$ depends on the relative orientation between the anisotropic direction, the direction of motion of the parton, and the direction of momentum broadening measurement. Their results indicate that the jet quenching parameter, $\hat{q}$ can be either smaller or larger than its isotropic value, depends sensitively on whether the comparison is made at equal temperatures or entropy densities. While studying the drag force in the anisotropic background shows that the results are not sensitive to whether the anisotropic and the isotropic plasmas are at equal temperatures or at equal entropy densities \cite{Chernicoff:2012iq}.

Recently, the anisotropic MT model has been studied at finite chemical potential \cite{Cheng:2014qia}. From the holographic point of view, the anisotropic plasma with chemical potential corresponds to an anisotropic charged black brane. The drag force acting on a massive quark moving through such medium has been studied in \cite{Cheng:2014fza,Chakraborty:2014kfa}. The effects of charge and anisotropy parameter on the jet quenching parameter were investigated in \cite{Wang:2016noh}.

Jets are studied in the context of gauge/gravity duality using different approaches \cite{Herzog:2006gh,Chesler:2008wd,Chesler:2008uy,Morad:2014xla, Ficnar:2013qxa,Ficnar:2013wba,Casalderrey-Solana:2015tas,Rougemont:2015wca,Rajagopal:2016uip,Brewer:2017dwd,Morad:2017wyx,Brewer:2017fqy}. In some studies, jets of light quark are described using the falling strings in the AdS background \cite{Herzog:2006gh,Chesler:2008wd,Chesler:2008uy,Morad:2014xla}. Some authors, approximate the motion of the string endpoint with a null geodesic and then study the dynamics of jet using the null geodesic equations in AdS-BH metric \cite{Chesler:2008wd,Chesler:2014jva,Chesler:2015nqz}. Using study a null geodesic in the WKB approximation, the jet quenching of a light probe in the anisotropic MT background has been studied in \cite{Muller:2012uu}. They find that the stopping distance of light quark is decreased by the anisotropic effect. However, it is shown that the stopping distances is slightly increased in the anisotropic plasma compared to its value in the isotropic plasma with the same entropy density once the quark is moving in the transverse direction.

On the other hand, pQCD calculations on jet energy loss show that the resulting medium-induced gluon radiation does not depend solely on the energy density of the medium, but also on the collective flow. Their results show that flow effects lead to a characteristic breaking of the rotational symmetry of the average jet energy \cite{Armesto:2004pt, Armesto:2004vz}. They also show that the partons lose less energy if they travel along the transverse direction in the collision.

Despite of the many efforts to understand the jets theoretically, different aspects of jet physics in the strongly coupled medium are still in need of better understanding. Using the gauge-string duality may help to understand unique aspects of showering and better modeling of them \cite{Rajagopal:2016uip}.

Our paper is organized as follows. In \sect{\ref{section:AnisoBG}} we give a brief review of the anisotropic MT background. The dynamics of the falling string is described in \sect{\ref{section:FallingStrings}}. One important result of studying the falling strings in the anisotropic background is understanding the stopping distance of the light quarks, presented in \sect{\ref{section:Stoppingdistance}}. Our simulation for an ensemble of jets, indicate that the stopping distance decreases significantly in the anisotropic background. We follow the evolution of the falling string in detail and study how the anisotropic parameter and the chemical potential affect the maximum stopping distance. We find that the maximum stopping distance is very sensitive to the anisotropy parameter and is smaller in the direction of beam. We close with Conclusions and Discussion in  \sect{\ref{section:Conclusions}}.

\section{Anisotropic background}
\label{section:AnisoBG}

In this section we review the anisotropic MT model which is the gravity dual to a static strongly coupled anisotropic plasma. The model is based on \cite{Azeyanagi:2009pr} with AdS boundary conditions. In the next subsection, we introduce the $U(1)$ charge in the background and study the anisotropic MT model at finite chemical potential \cite{Cheng:2014qia}.

The anisotropy is introduced by deformation of $\mathcal{N}=4$ SYM by adding a $\theta$ term to the action that depends linearly on one of the three spatial coordinates $z$ as $ \theta = 2\pi\,n_{D7}\,z$, where $n_{D7}$ can be thought as the density of D7-branes homogeneously distributed along the anisotropic direction. Therefore, the action is
\bea
S=S_{\mathcal{N}=4}+\frac{1}{8\pi^2}\int \theta(z) Tr\left(F  F\right).
\eea

\begin{table}
	\begin{center}
		\begin{tabular}{cc||cccc|c|c}
			& & $t$ & $x$ & $y$ & $z$ & $u$ & $S^5$ \\
			\hline
			$N_c$ & $D3$ & x & x & x & x &- &- \\
			$n_{D7}$ & $D7$ & x & x & x &- &- & x \\
				\end{tabular}
	\end{center}
	\caption{D3D7 set up.}
	\label{tab:brane}
\end{table}

The type IIB supergravity solution is static and anisotropic but at equilibrium with finite temperature. Moreover it is regular on and outside the horizon and asymptotically approaches AdS$_5$. The brane set up is given by Tabel. 1. There is no flavor degrees of freedom with such embedding of D7 brane.

The AdS part of the supergravity solution in the string frame is given by \cite{Mateos:2011ix}
\bea
&&ds^2=\frac{L^2}{u^2}\left(-\mathcal{F}\mathcal{B}dt^2+dx^2+dy^2+\mathcal{H}dz^2+\frac{du^2}{\mathcal{F}}\right)\nn\\
&&\phi=\phi(u),\hspace{10mm}\chi=a\,z,     \label{aniso-background}
\eea
where $\phi$ and $\chi$, are the dilaton and axion fields respectively, and  the anisotropy parameter is  $a=\frac{g_{YM}^2n_{D7}}{4\pi}$. Also the AdS radius is $L$.
The metric functions are $\mathcal{F}, \mathcal{B}$ and $\mathcal{H}$ which depend on the holographic radial coordinate $u$, the radius of horizon, $u_H$ and the anisotropy parameter, $a$. The boundary of space is at $u=0$ and the horizon located at $u=u_H$ where the blackening function is zero, $\mathcal{F}(u_H)=0$. The isotropic solution is given by $a=0$ and
\bea
\phi=0,\,\,\,\,\ \mathcal{F}=1,\,\,\,\,\ \mathcal{B}=\mathcal{H}=1.
\eea

The temperature is given by
\bea
T=\frac{1}{4\pi}\mathcal{F}'(u_h)\sqrt{\mathcal{B}(u_H)}.
\eea
The entropy density equals a quarter of the horizon area over the spatial volume as
\bea
s=\left(\frac{\pi^2N_c^2}{2}\right)\frac{e^{\frac{-5}{4}}\phi_H}{\pi^3u_H^3}.
\eea
The solutions of this background are parameterized by the dilaton field value at the horizon, $\phi_h$ and the horizon, $u_H$. From the field theory point of view, they map to the anisotropy parameter, $a$, and the temperature of the plasma, $T$. Two functions for the values of $u_H$ and $\phi_h$ in terms of $a$ and $T$ are obtained in \cite{Ali-Akbari:2014nua}.

One should consider two different anisotropic plasmas to compare the anisotropic results to that in the isotropic case.  In the first case, two plasmas can be taken to have the same temperature, $T$ but with different entropy density $s$. In the second case, the entropy densities are the same but the temperatures differ. We will consider both cases and compare the final results with isotropic plasma. \footnote{In the numeric calculations, we use $\tilde{\phi}_h=\phi_H+4/7 \log{a}.$}\\

At small $a/T$, the entropy density scales as in the isotropic case $S_{iso}=\frac{\pi^2}{2}N_c^2 T^3$,
while at large $a/T$ it scales as $s = c_{ent}N^2_c a^{1/3} T^{8/3}$, where $c_{ent}$ is a constant. For small $a/T$ the metric functions and the radius of the horizon are known as some expansions around the black D3-brane solution \cite{Mateos:2011ix},
\bea
&&\mathcal{F}=1-\frac{u^4}{u_H^4}+a^2\mathcal{F}_2+\mathcal{O}(a^4),\nn\\
&&\mathcal{B}=1+a^2\mathcal{B}_2+\mathcal{O}(a^4),\nn\\
&&\log\mathcal{H}=\frac{a^2u_H^2}{4}\log\left(1+\frac{u^2}{u_H^2}\right)+\mathcal{O}(a^4),\nn\\
&&u_H=\frac{1}{\pi T}+\frac{5\log 2-2}{48\pi^3T^3}a^2+\mathcal{O}(a^4),\label{small-a-T}
\eea
in which
\bea
&&\mathcal{F}_2=\frac{1}{24u_H^2}\left[8u^2(u_H^2-u^2)-10u^4\log2+3u_H^4+7\log\left(1+\frac{u^2}{u_H^2}\right)\right],\\
&&\mathcal{B}_2=-\frac{u_H^2}{24}\left[\frac{10u^2}{u_H^2+u^2}+\log\left(1+\frac{u^2}{u_H^2}\right)\right].
\eea

The pressures and energies which are useful on making a naive connection with the weak coupling results can be found from the expectation values of the stress energy tensor and read
\bea
\label{ppxz} &&E= 3 e_1+ a^2 e_2+{O}(a^4)~,\quad P_{\perp}= e_1+ a^2
e_2+{O}(a^4)~,\quad P_{\parallel}= e_1- a^2 e_2+{O}(a^4)~,
\eea
where $ e_1=\frac{ N_c^2  T^2}{32}+{O}(a^4)~,
e_2=\frac{ N_c^2  T^2}{32}$ and $N_c$ is the number of colors.
Notice that \be\label{ppp1} P_\parallel<P_\perp. \ee

\subsection{Anisotropic background at finite chemical potential}
\label{subsection:ChemicalP}

Because of non-zero chemical potential in the QGP, the anisotropic MT background in the presence of finite chemical potential has been studied in \cite{Cheng:2014qia}. The extension has been done by considering a charged black brane in the background to obtain the anisotropic charged AdS black hole solution.

The AdS part of the the solution is given by \eq{aniso-background} together with $A_t=A_t(u)$. Therefore, the corresponding chemical potential $\mu$ is given by
\be
\mu=\int_0^{u_H}Q\sqrt{\cB}e^{3/4\phi}u\,du.
\ee
Here, $Q$ is related to the charge density. The metric functions express in terms of the dilaton as follows
\bea
\cH&=& e^{-\phi} , \label{eq_H} \\
\cF&=&\frac{e^{-\frac{1}{2}\phi}}{12(\phi'+u\phi'')}\left(3a^2 e^{\frac{7}{2}\phi}(4u+u^2\phi')+48\phi'-2e^{\frac{5}{2}\phi}Q^2u^6\phi'\right),
\label{eq_F} \\
\frac{\cB'}{\cB}&=&\frac{1}{24+10 u\phi'}\left(24\phi'-9u\phi'^2+20u\phi''\right)\,\label{eq_B},
\eea
where the prime denotes derivative with respect to $u$. The dilaton itself should satisfy a third-order equation
\begin{eqnarray}
0&=&\frac{-48 \phi '^2 \left(32+7 u \phi '\right)+768 \phi ''+4 e^{\frac{5 \phi }{2}} Q^2 u^5 \left(-24 \phi '+u^2 \phi
	'^3-8 u \phi ''\right)}{48 \phi '-2 e^{\frac{5 \phi}{2}} Q^2 u^6 \phi '+3 a^2 e^{\frac{7 \phi}{2}} u \left(4+u \phi '\right)}
+\frac{1}{u \left(12+5 u \phi '\right) \left(\phi '+u \phi ''\right)}\nonumber\\
&& \times\Big[13u^3\phi'^4+u^2\phi'^3(96+13u^2\phi'')+8u(-60\phi''+11u^2\phi''^2-12u\phi^{(3)}) \nonumber\\
&&~~~~~+2u\phi'^2(36+53u^2\phi''-5u^3\phi^{(3)})+\phi'(30u^4\phi''^2-64u^3\phi^{(3)}-288+32u^2\phi'')\Big].
\label{eq_dil}
\end{eqnarray}
As one expects, the asymptotical $AdS_5$ boundary conditions require that $\phi(0)=0$,  $\cF(0)=\cB(0)=1$, and then $\cH(0)=1$.
To solve these differential equations, one should consider
\be
\phi \rightarrow \tilde{\phi}=\phi+\frac{4}{7}\log a,\,\,\,\,\,Q=a^{\frac{-5}{7}}Q.
\ee
In this way, one eliminates $a$ from the differential equations and boundary conditions for $\phi(u)$. Then one obtains the anisotropic charged black brane solution, numerically. It should be noticed that two class of solutions are oblate $(a^2>0)$ and prolate $(a^2<0)$ solutions. We consider the physical class where $(a^2>0)$.

\section{Falling strings in the anisotropic background}
\label{section:FallingStrings}

\begin{figure}
\center
\includegraphics[scale=0.55]{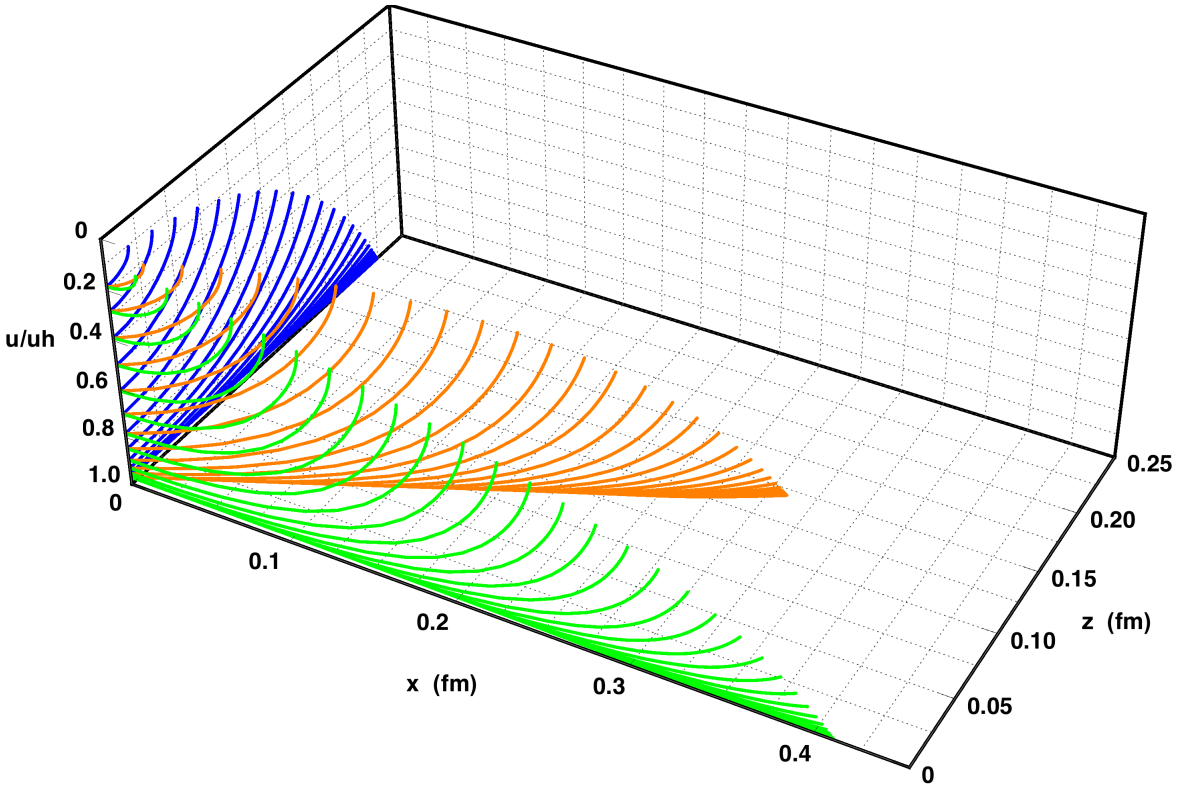}
\caption {(Color online) 
The typical falling string profiles moving in the $x-z$ plane in the anisotropic background, with the angle $\phi=0$ (blue),  $\phi=\pi/3 $ (orange), and $\phi=\pi/2 $ (green) where $\phi$ is measured with respect to the z-axis. Each line shows the string at different instant time.  The strings are created at a point at $u_c = 0.1~u_h$ and evolve to the extended objects by time evolution.  The value of the anisotropy parameter is $a=4.4$. The quark, corresponding to the half of the string, has $100$ GeV energy and the temperature of the plasma is $350$ MeV. 
\label{string}}
\end{figure}

Now, we study how the anisotropy and the chemical potential affect the light quark jet dynamics. The physical setup of interest is one of a back-to-back jet pair created in the anisotropic QGP. We therefore consider configurations for which the string is created at a point and expands in space-time such that the two endpoints of the string move away from each other. One should notice that the total spatial momentum of the string vanishes. Due to the rotational symmetry in the transverse plane $(x-y)$, we consider the moving end points only in the $(x-z)$ plane with an angle $\phi$ with respect to the beam direction $z$ coordinate. In this case, the jet direction along the beam direction and transverse directions correspond to $\phi=0$ and $\phi=\pi/2$, respectively. Therefore, the embedding function of string $X^{\mu}(\tau,\sigma)$ will be a map to $( t(\tau,\sigma),x(\tau,\sigma),0,z(\tau,\sigma),u(\tau,\sigma))$.\\

In \fig{string}, one finds a numerically generated falling open string in the $x-z$ plane in the x direction (Green), with a angle $\phi=\pi/3$ with respect to the z axis (Orange), and in the z direction (Blue), obtained numerically in the anisotropic background \refeq{aniso-background}. In this figure, the open string has $200$ GeV energy and the temperature of the plasma is $350$ MeV. The initial condition has been chose such that at zero worldsheet time, the open string is mapped to a single point. Such open string profile created at time $t=t_c$ is given by

\begin{equation}
    t(0,\sigma)=t_c\, ,\,\,\, x(0,\sigma)=0\, ,\,\,\,z(0,\sigma)=0\, ,\,\,\,u(0,\sigma)=u_c,
\end{equation}
where $\sigma\in[0,\,\pi]$.  After the creation at time $t_c$, the open string expands from a point into an extended object while the open string endpoints fall toward the horizon. One concludes that in the anisotropic background the extension of the string in the $x$ direction differs from its extension in the $z$ direction, significantly.\\

Now we explain the numerical procedure to produce the profile of falling strings in \fig{string}. The dynamics of the open string in the anisotropic background is governed by the Nambu-Goto action which is classically equivalent to the Polyakov action.  However, it is found that the numerical studies of the string profile is more convenient by using the Polyakov action, \cite{Herzog:2006gh,Chesler:2008wd,Chesler:2008uy}. The Polyakov action for the string has the form
\begin{equation} \label{Polyakov action}
    S_P=-\frac{T_0}{2}\int d^2\sigma \> \sqrt{-\eta} \, \eta^{ab}\,
    \partial_a X^\mu\partial_b X^\nu \, G_{\mu\nu}\, .
\end{equation}
By variation of the Polyakov action with respect to the embedding functions $X^\mu$, one finds the equation of motion as
\bea
	\label{stringEoM}
	\partial_{a} \big[ \sqrt{-\eta}\,\eta^{ab} \, G_{\mu\nu}\, \partial_{b}X^{\nu}  \big] & = \frac{1}{2} \sqrt{-\eta}\,\eta^{ab}
    \frac{\partial G_{\nu \rho}}{\partial X^{\mu}} \,
    \partial_{a}X^{\nu}\partial_{b}X^{\rho} \nonumber \\[5pt]
    \Longleftrightarrow \qquad \nabla_a\,  \Pi^a_{\mu} & = -\frac{T_0}{2} \,\eta^{ab}
    \frac{\partial G_{\nu \rho}}{\partial X^{\mu}} \,
    \partial_{a}X^{\nu}\partial_{b}X^{\rho},
\eea
where $\Pi^a_\mu$ are the canonical momentum densities associated with the string and are obtained from varying the action with respect to the derivatives of the embedding functions. The worldsheet metric is chosen as
\begin{equation} \label{Worldsheet Metric}
    \|\eta_{ab}\|= \left(\begin{array}{cc} -\Sigma(x,u) & 0 \\ 0 & \Sigma(x,u)^{-1}
    \end{array}\right),
\end{equation}
where stretching function, $\Sigma(x,u)$ can be any arbitrary function of $x$ and $u$. The point is that we need to choose this function such that the singularities in the equations of motion are cancelled. We found that the stretching function as
\begin{equation}
\Sigma(x,u)=
\left(\frac{1-u/u_H}{1-u_{\rm c}/u_H}\right)^b \left(\frac{u_{\rm c}}{u}\right)^c
 \,\label{eq:stretch} ,
\end{equation}
will keep the equations of motion well behaved in the anisotropic background as well. We choose the values of $b$ and $c$ case by case in the range of 1 to 3.

In order to solve the equations of motion in the anisotropic background we need a self-consistent initial conditions (IC) for the string profile which obey the constraint equation,
\begin{equation}
    -\mathcal{F}(u)^2 \mathcal{B}\, \dot t^2+\mathcal{F}(u)\, \dot x^2+\mathcal{H}\mathcal{F}\, \dot z^2+\dot u^2=0
    \,\label{con3} .
\end{equation}
The $\sigma$ derivatives of $X^\mu$ are initially zero for the string with point-like IC. We choose the following IC for the string in this background
\begin{eqnarray}
   &&  \dot{x}(0,\sigma) =A  \,u_c \cos (\sigma ) \sin (\phi ) \,, \nonumber \\
   &&  \dot{z}(0,\sigma) =A  \,u_c \cos (\sigma ) \cos (\phi ) \,, \nonumber \\
    && \dot{u}(0,\sigma) = u_{\rm c}\sqrt{f(u_{\rm c})} \, (1-\cos 2\sigma)\,, \\
   && \dot{t}(0,\sigma) = u_c \sqrt{\frac{A^2 \cos ^2(\sigma ) \left(H(u_c) \cos ^2(\phi )+\sin ^2(\phi )\right)+(1-\cos (2 \sigma ))^2}{B(u_c)}}    \, ,\label{eq:IC}  \nonumber
 \end{eqnarray}
where $u_c$ and $A$ are free parameters that can be related to the energy and momentum of the dual light quark in the field theory (see the next subsection). These IC yield a string profile that is symmetric about $r=0$ at all times, because $\dot{x}(0,\sigma)$ and $\dot{z}(0,\sigma)$ are antisymmetric about $\sigma=\pi/2$ while $\dot{u}(0,\sigma)$ is symmetric.

To study different properties of falling strings in the anisotropic background, one should solve the equation of motion Eq.~(\ref{stringEoM}) numerically with \emph{Mathematica}'s \texttt{NDSolve} to obtain the embedding functions $X^{\mu}$ as a function of $(\tau,\sigma)$.

The solution of Eq.~(\ref{stringEoM}) in the anisotropic background for the values of $\phi$ is plotted in \fig{string} . In order to be more clear, we just plot the half of the string. The dynamics of the other half of the string is exactly the same, but moving in the opposite direction. As we expected, the two endpoints of the string move away from each other as the string extends along the $(x, \, z)$ direction and falls toward the horizon. Numerical results show that the string in the beam direction is suppressed with respect to the transverse direction even at small anisotropy parameter.

\subsection{Definition of the quark virtuality}
\label{subsection:virtuality}

It is important to make a minor comment on finding the the energy and momentum of the light quark from the falling open string.
The anisotropic background geometry $G^{\mu\nu}$ depends only on $u$, hence for $\mu$ corresponding to $(t,\,\vec{x})$ one finds
\begin{equation}
\nabla_a\Pi_\mu^a = 0. \label{covariant EQM}
\end{equation}
Therefore, the corresponding momentum densities $\Pi^a_\mu$ are conserved Noether currents on the related worldsheet.

The flow of the $\mu$ component of the space-time momentum of the open string along the $a$ direction on the worldsheet is given by $\Pi_\mu^a$ \cite{Lawrence:1993sg}. So the total energy of the falling string in the anisotropic geometry Eq~.(\ref{aniso-background}) at any time is constant and equal to the initial energy of the string
\begin{eqnarray}
E_{\rm string}  &=& - \int_0^\pi d \sigma \> \sqrt{-\eta} \, \Pi^\tau_{t}(0,\sigma) \nonumber \\
&=& \frac{\lambda}{2 \pi} \int_0^\pi d \sigma \> \frac{\mathcal{F}(u_c)}{u_c\, \Sigma(u_c)}\, \sqrt{\frac{A^2 \cos ^2(\sigma ) \left(\mathcal{H}(u_c) \cos ^2(\phi )+\sin ^2(\phi )\right)+(1-\cos (2 \sigma ))^2}{\mathcal{B}(u_c)}} \, .   \nonumber \\
\end{eqnarray}
Then the total energy of light quark is given by
\begin{equation}
E_q=\frac{1}{2} E_{\rm string}.
\end{equation}
Also, the total momentum in $x$ and $z$ direction is conserved and can be calculated by the string IC as follows:
\begin{equation}
P_{q\>x} = \frac{\sqrt \lambda}{2\pi}\frac{A\, \sin(\phi)}{u_c\,\Sigma(u_{\rm c})} \,\,\,\,\,\, , \,\,\,\,\,\,P_{q\>z} = \frac{\sqrt \lambda}{2\pi}\frac{A\, \mathcal{H}(u_c)\,\cos(\phi)}{u_c\,\Sigma(u_{\rm c})}.
\end{equation}
In the above integral, we used the fact that the total momentum of the string in $x-z$ direction is zero and it is completely symmetric about the point $\sigma=\pi/2$. So, the momentum of the quark is equal to the momentum of the anti-quark in the opposite direction. We can see that the two parameters $A$ and $u_c$ in the IC of the string determine the energy and momentum of the string completely.

Now, the virtuality of the corresponding jet in the anisotropic QGP is defined as
\begin{equation}
	Q^2 \equiv E_q^2 - P_q^2 \,,
\end{equation}
where $P_q^2=P_{q\>x}^2+P_{q\>z}^2$ is the total spatial momentum of the quark in the $x-z$ plane.

\section{Stopping distance of falling string in the anisotropic background}
\label{section:Stoppingdistance}

The maximum stopping distance of the falling string can be used as a phenomenological guideline to estimate the stopping power of the strongly- coupled plasma.  This distance which is called the thermalization distance, $x_{therm}$, is defined as the length along the x direction from the point of production of the original point-like string to the point at which the end of the string falls through the black hole horizon. On the field theory side of the duality, $x_{therm}$ corresponds to the length of the plasma traversed before the jet becomes completely thermalized (i.e. indistinguishable from the plasma). However, in general this information alone is not enough to calculate observables such as $R_{AA}$ or $v_2$.

\begin{figure}
\center
\includegraphics[scale=0.65]{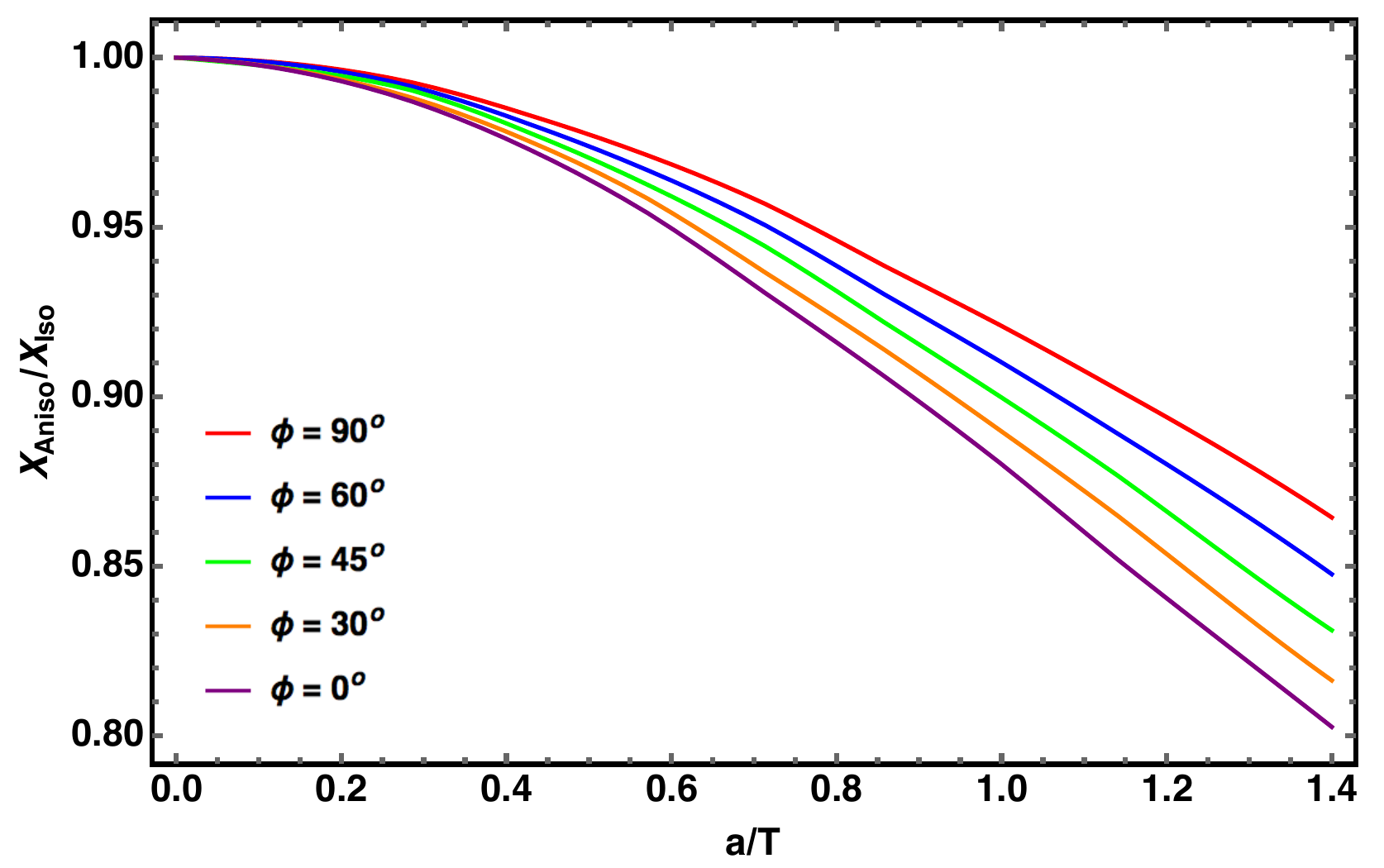}
\caption {
The ratio of the stopping distances in the anisotropic geometry to the $AdS-Sch$ metric for different values of $a/T$. Red line shows the ratio for $x_{Aniso}=x_{transverse}$, while the purple line stands for $x_{Aniso}=x_{longitudinal}$. Here we fix the temperature of both backgrounds and the string has the same energy and virtuality in both backgrounds.
 \label{fig2}}
\end{figure}

In \fig{fig2}, we compare the stopping distance of a light quark moving in the anisotropic QGP with the same light quark moving in an isotropic QGP. In order to compare the anisotropic effect, we fix the temperature of the both backgrounds at $350$ MeV and the energy of the moving quark in both background at $100$ GeV. We find that even at the small anisotropy, $a/T <1$ the stopping distance in the anisotropic plasma is smaller than the isotropic plasma even when the string is moving in the transverse direction.

\begin{figure}[!htbp]
\centering
\begin{subfigure}[b]{.05in}
    \captionsetup{skip=-15pt,margin=-10pt}
    \caption{}
    \label{Rxa}
\end{subfigure}
\begin{subfigure}[b]{2.7in}
    \centering
     \includegraphics[width=2.7in]{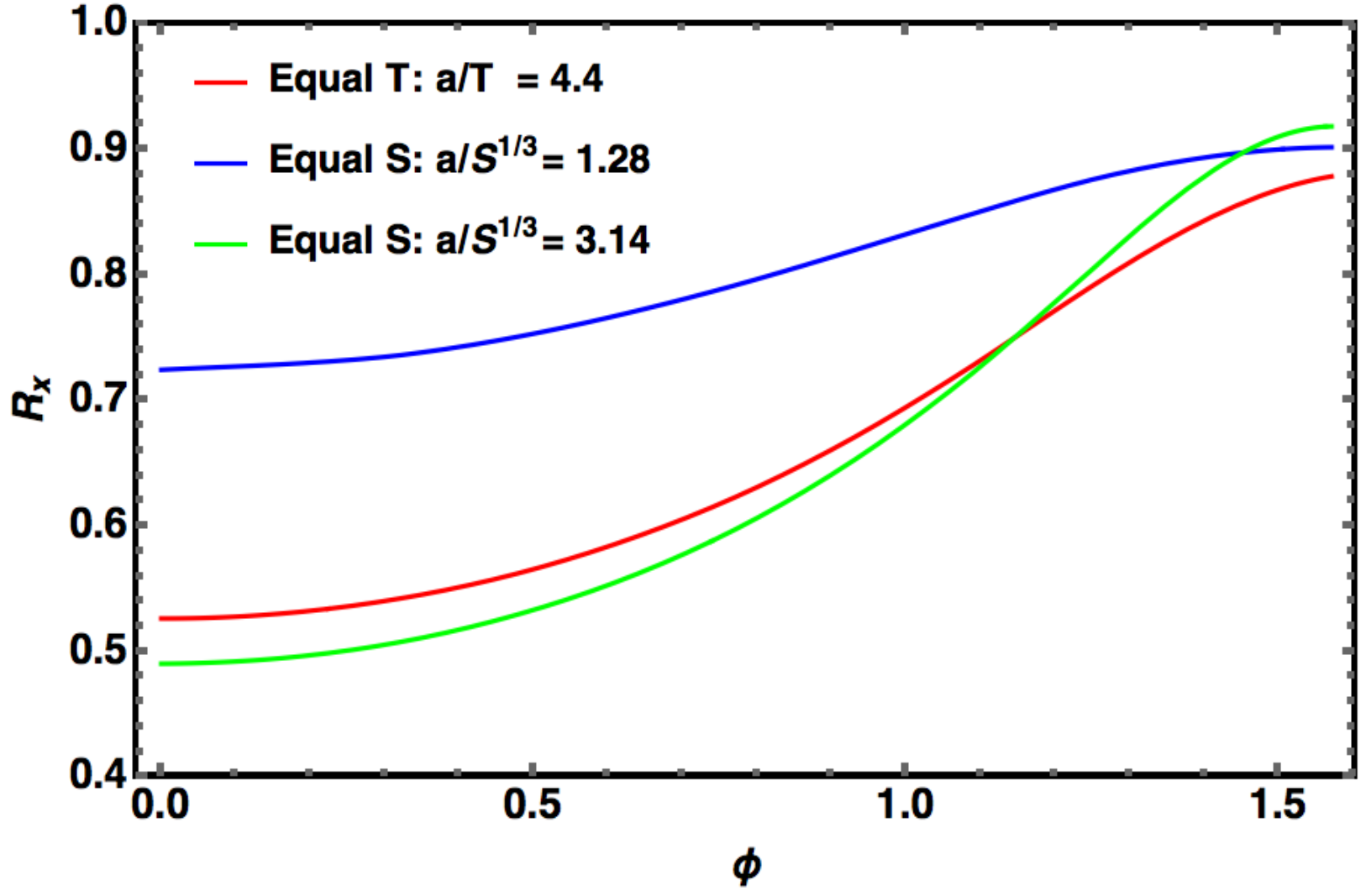}
\end{subfigure}
\begin{subfigure}[b]{.05in}
    \captionsetup{skip=-15pt,margin=-20pt}
    \caption{}
    \label{Rxb}
\end{subfigure}
\begin{subfigure}[b]{2.7in}
    \centering
    \includegraphics[width=2.7in]{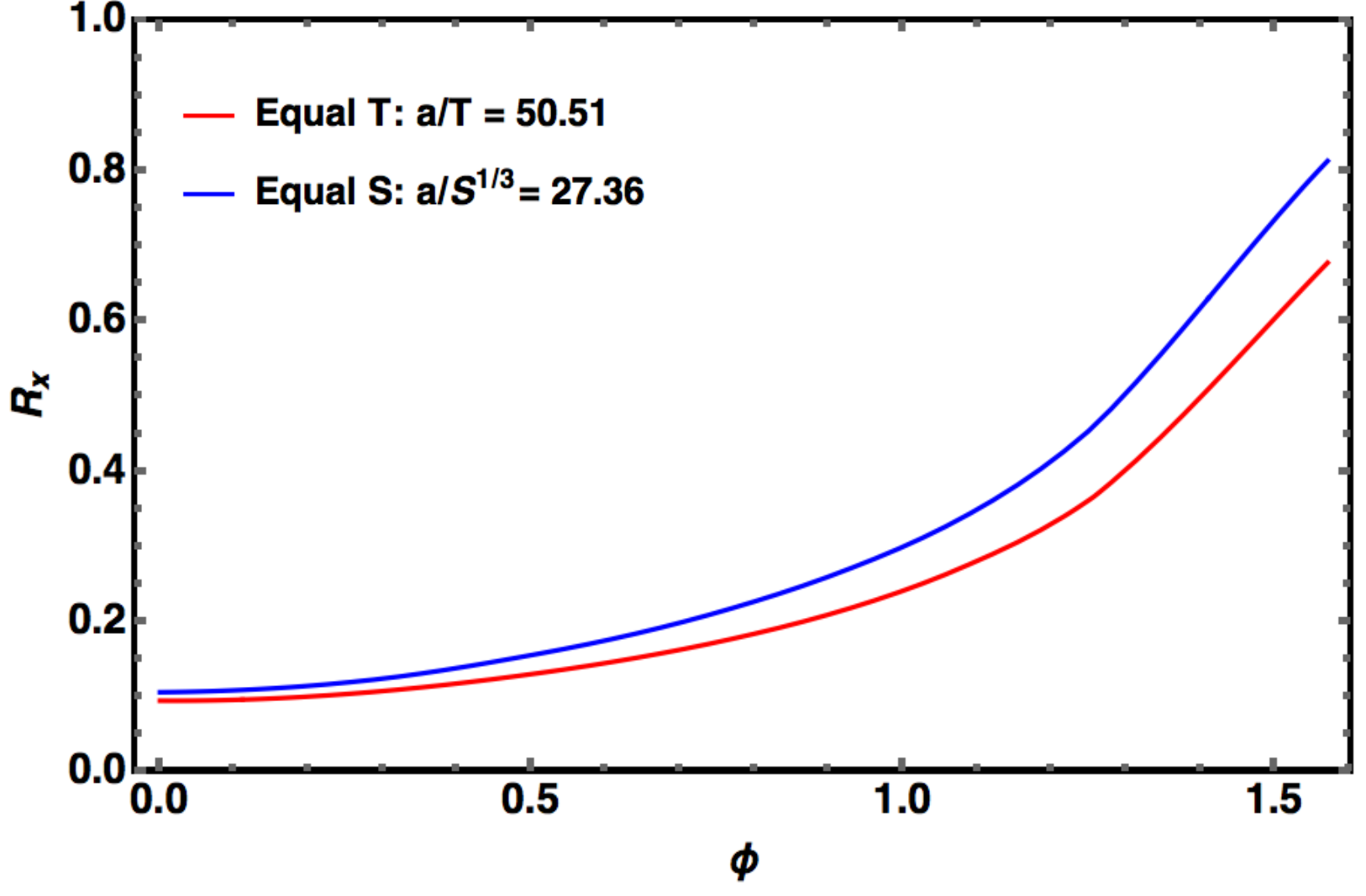}
\end{subfigure}
\qquad \qquad \qquad \qquad \qquad
\vspace{.05in}
\caption{(Color online)  (\subref{Rxa}) The ratio of stopping distance of the quark in the anisotropic medium to the stopping distance in the isotropic medium $R_x=x_{aniso}/x_{iso}$ at mid anisotropy at equal temperature (red) and equal entropy density (blue, green). Here we take $P_q\simeq 0.9913 \, E_q$. (\subref{Rxb}) The ratio $R_x=x_{aniso}/x_{iso}$ at large anisotropy at equal temperature (red) and equal entropy density (blue).}
\label{Rx}
\end{figure}

It is interesting to see how the direction of motion affects the stopping distance of the string. In \figtwo{Rx}{Rxa}, we plot the ratio of thermalization distance of the string in the anisotropic medium to the thermalization distance of the same string in the AdS-Sch background. Our results at mid-anisotropy show that the suppression of the stopping distance becomes less prominent at equal entropy density. The \figtwo{Rx}{Rxb} Shows the same qualitative results at the medium with larger anisotropy, although supression of string is larger as we expected from the previous plot. In these plots, we fix the ratio of the quark's energy to its momentum such that $P_q \simeq 0.9913 \,E_q$.

\begin{figure}[!htbp]
\centering
\begin{subfigure}[b]{.05in}
    \captionsetup{skip=-15pt,margin=-10pt}
    \caption{}
    \label{XEa}
\end{subfigure}
\begin{subfigure}[b]{2.8in}
    \centering
     \includegraphics[width=2.8in]{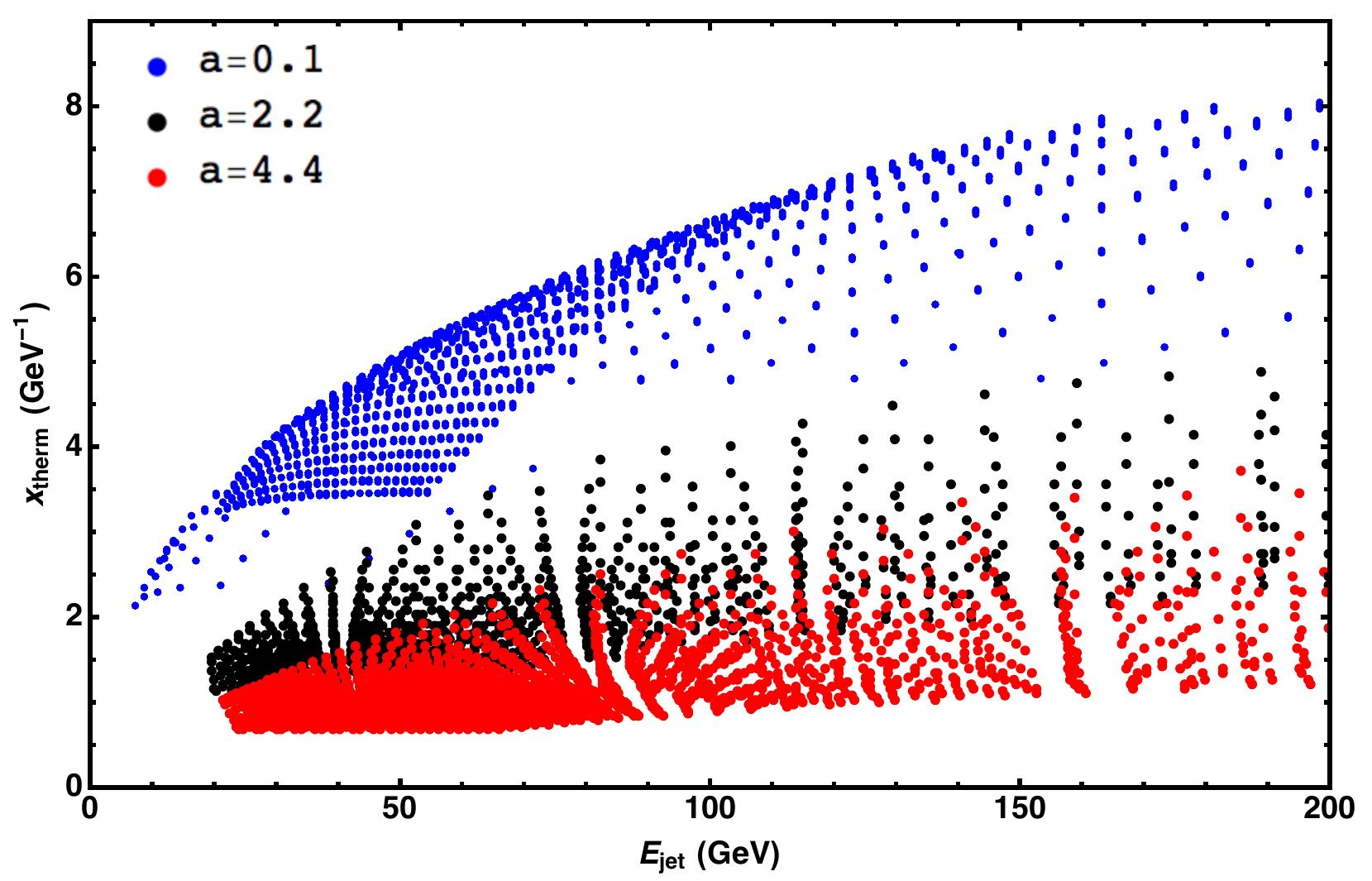}
\end{subfigure}
\begin{subfigure}[b]{.05in}
    \captionsetup{skip=-15pt,margin=-20pt}
    \caption{}
    \label{XEb}
\end{subfigure}
\begin{subfigure}[b]{2.8in}
    \centering
    \includegraphics[width=2.8in]{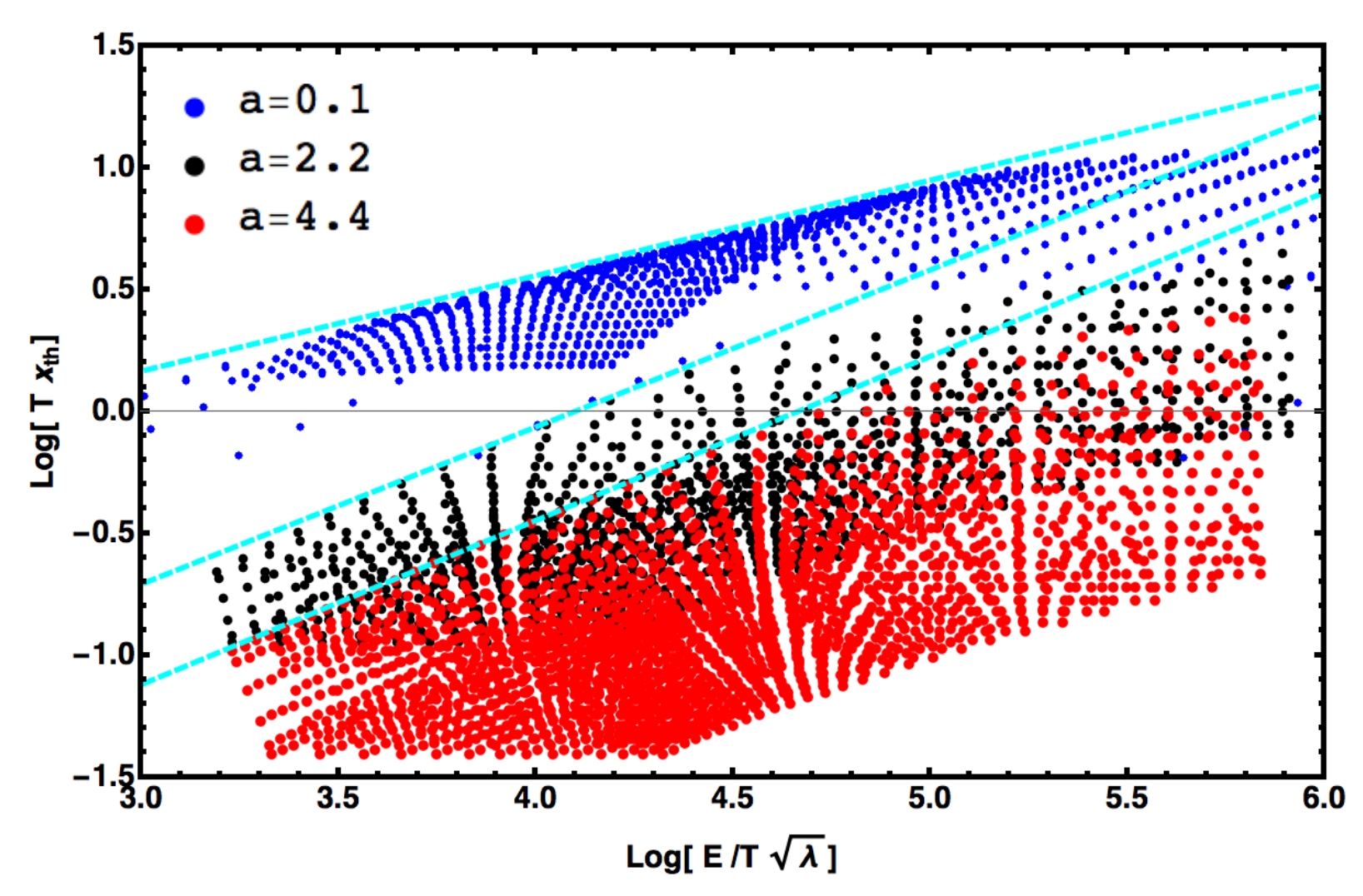}
\end{subfigure}
\vspace{.05in}
\caption{(Color online)  (\subref{XEa}) The stopping distance of the jet moving in the anisotropic backgrounds with different anisotropy parameters in terms of its energy and (\subref{XEb}) The log-log plot of the quark stopping distance as a function of total quark energy E for many falling strings with different initial conditions. For each sets of data, all data points fall below the blue line given by $x_{therm}=(\mathcal{C}/T)\,(E/T\,\sqrt{\lambda})^{n_{eff}}$. Our results show that the $n_{eff}=0.376,\,0.633\,$and $0.666$ for $a=0.1\,,2.2$ and $4.4$.}
\label{X-E}
\end{figure}

We find that the maximum stopping distance of the quark depends sensitively on the IC of the falling string \cite{Morad:2014xla}. However, numerical results show that the maximum stopping distance for a given energy depends on the energy of the quark and temperature of the plasma as follows \cite{Chesler:2008uy}
\begin{equation}
	x_{therm}=\frac{\mathcal{C}}{T} \,(\frac{E}{T\, \sqrt{\lambda}})^{1/3} \,,\label{XE}
\end{equation}
where the constant $\mathcal{C}=0.526$ is obtained using the numerical calculations in $AdS-Sch$ background \cite{Chesler:2008uy}. We numerically compute the maximum stopping distance for strings with different initial conditions in anisotropic medium with different anisotropy parameters. Our results are illustrated in \figtwo{X-E}{XEa}. Each point shows the dynamics of an string with specific initial conditions. Blue dots show the results in the medium with a small anisotropy parameter $a=0.1$ while the black and red dots are results calculated in the medium with anisotropy parameter $a=2.2$ and $a=4.4$, respectively. In \figtwo{X-E}{XEb} all data points for each sets of data fall below the blue line given by $x_{therm}=(\mathcal{C}/T)\,(E/T\,\sqrt{\lambda})^{n_{eff}}$. Our results show that the $n_{eff}=0.376$ for $a=0.1$. While by increasing the anisotropy of the medium, this effective power is also increased such that $n_{eff}=0.633$ and $n_{eff}=0.666$ for $a=2.2$ and $a=4.4$, respectively.

In \cite{Ficnar:2011yj}, the stopping distance of the light quark in the non-conformal background is plotted in terms of the quark energy and it is shown that the effective power in Eq. (\ref{XE}) becomes temperature and energy-dependent. Their results indicate that this power is much larger than $1/3$ at small energies $(\sim 100 GeV)$. Although, this power reduces to $1/3$ for enough large energies which is in agreement with our results.

\subsection{Anisotropic background with zero charge}
\label{subsection:StoppingdistancezeroP}%

In order to have a better understanding of the jet quenching of a light probe in the anisotropic plasma, we plot the maximum stopping distance for $E_q$ = 100 GeV quarks in the anisotropic plasma as a function of the creation position of the string in the radial direction $u_c/u_h$, which is equivalent to varying the virtuality $Q^2$ of the jet.
Since we wish to compare our results with the isotropic $AdS-Sch$ results, we consider the anisotropic plasma have either the same temperature or the same entropy density as the isotropic plasma.

\begin{figure}[!htbp]
\centering
\begin{subfigure}[b]{.05in}
    \captionsetup{skip=-15pt,margin=-10pt}
    \caption{}
    \label{ConstantTa}
\end{subfigure}
\begin{subfigure}[b]{2.8in}
    \centering
     \includegraphics[width=2.8in]{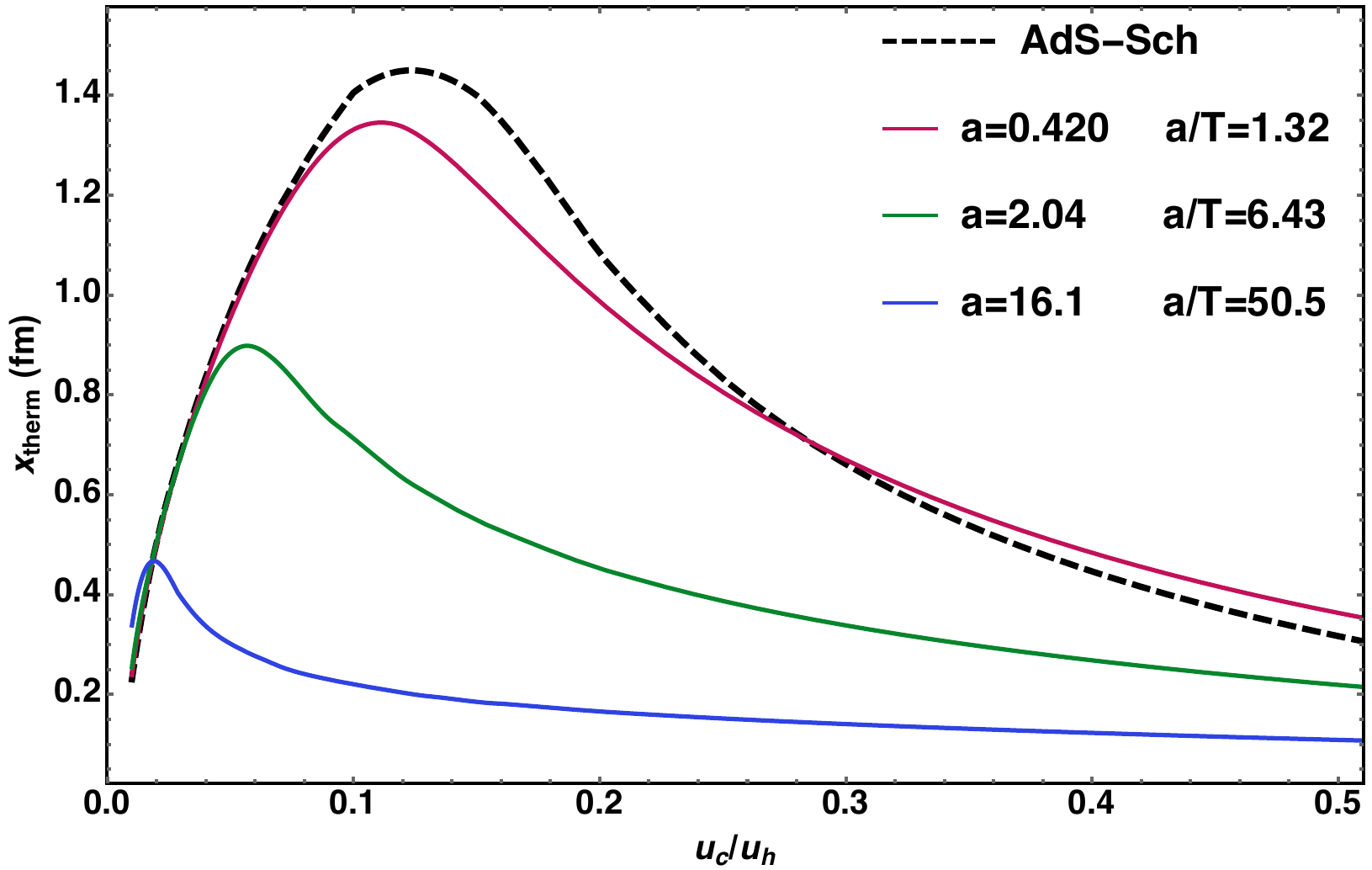}
\end{subfigure}
\begin{subfigure}[b]{.05in}
    \captionsetup{skip=-15pt,margin=-20pt}
    \caption{}
    \label{ConstantTb}
\end{subfigure}
\begin{subfigure}[b]{2.8in}
    \centering
    \includegraphics[width=2.8in]{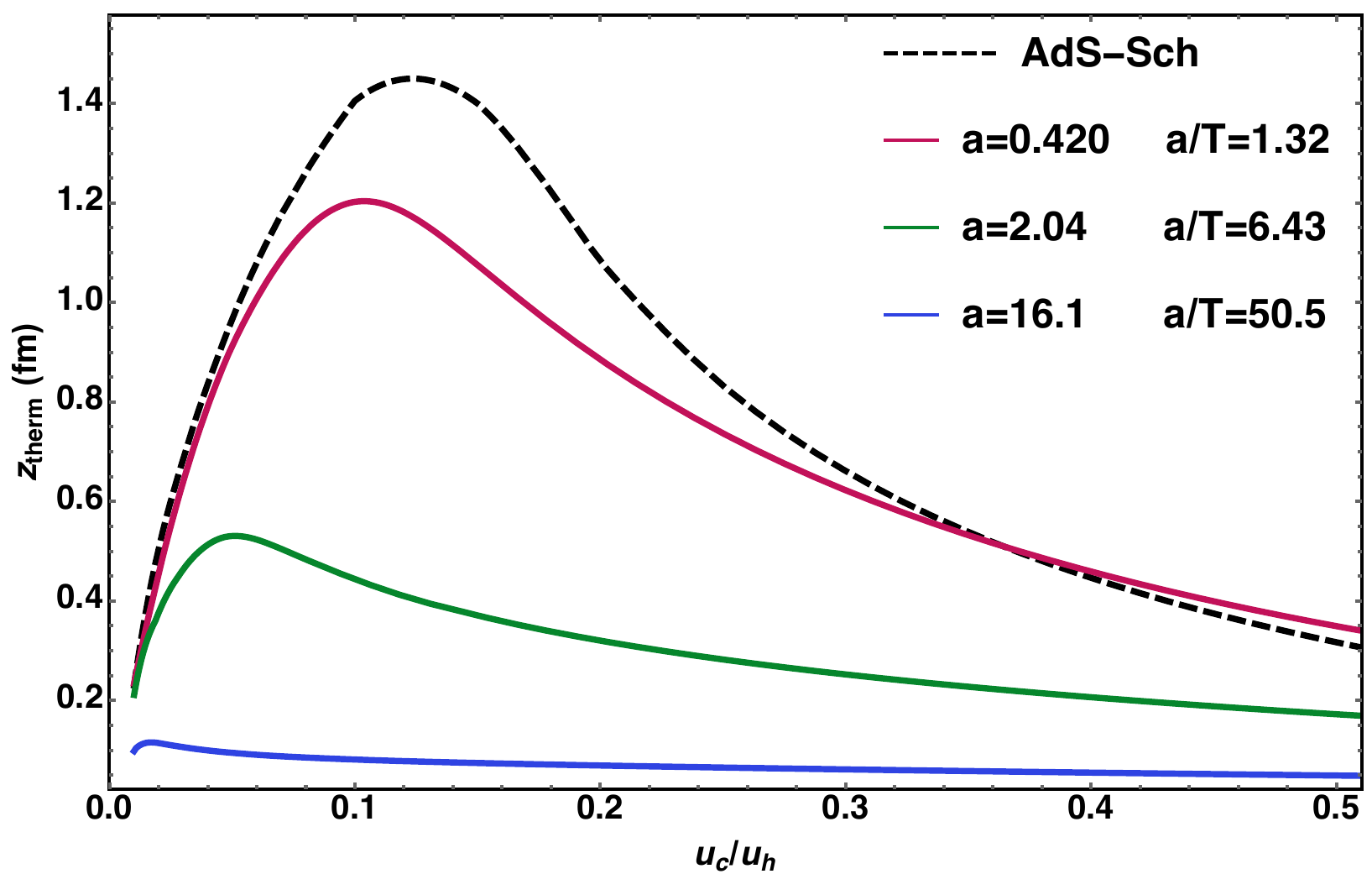}
\end{subfigure}
\vspace{.05in}
\caption{(Color online) The maximum stopping distance of the jet moving (\subref{ConstantTa}) in the x-direction and (\subref{ConstantTb}) in the z-direction in the anisotropic background with different anisotropy parameters. The anisotropic plasma has the same temperature as the isotropic plasma, $T=$ 318 MeV. In all cases, the quark has 100 GeV energy, but created at different radial distance from the boundary. The black dashed line is the stopping distance in the AdS-Sch background. }
\label{ConstantT}
\end{figure}

\begin{figure}[!htbp]
\centering
\begin{subfigure}[b]{.05in}
    \captionsetup{skip=-15pt,margin=-10pt}
    \caption{}
    \label{ConstantSa}
\end{subfigure}
\begin{subfigure}[b]{2.8in}
    \centering
     \includegraphics[width=2.8in]{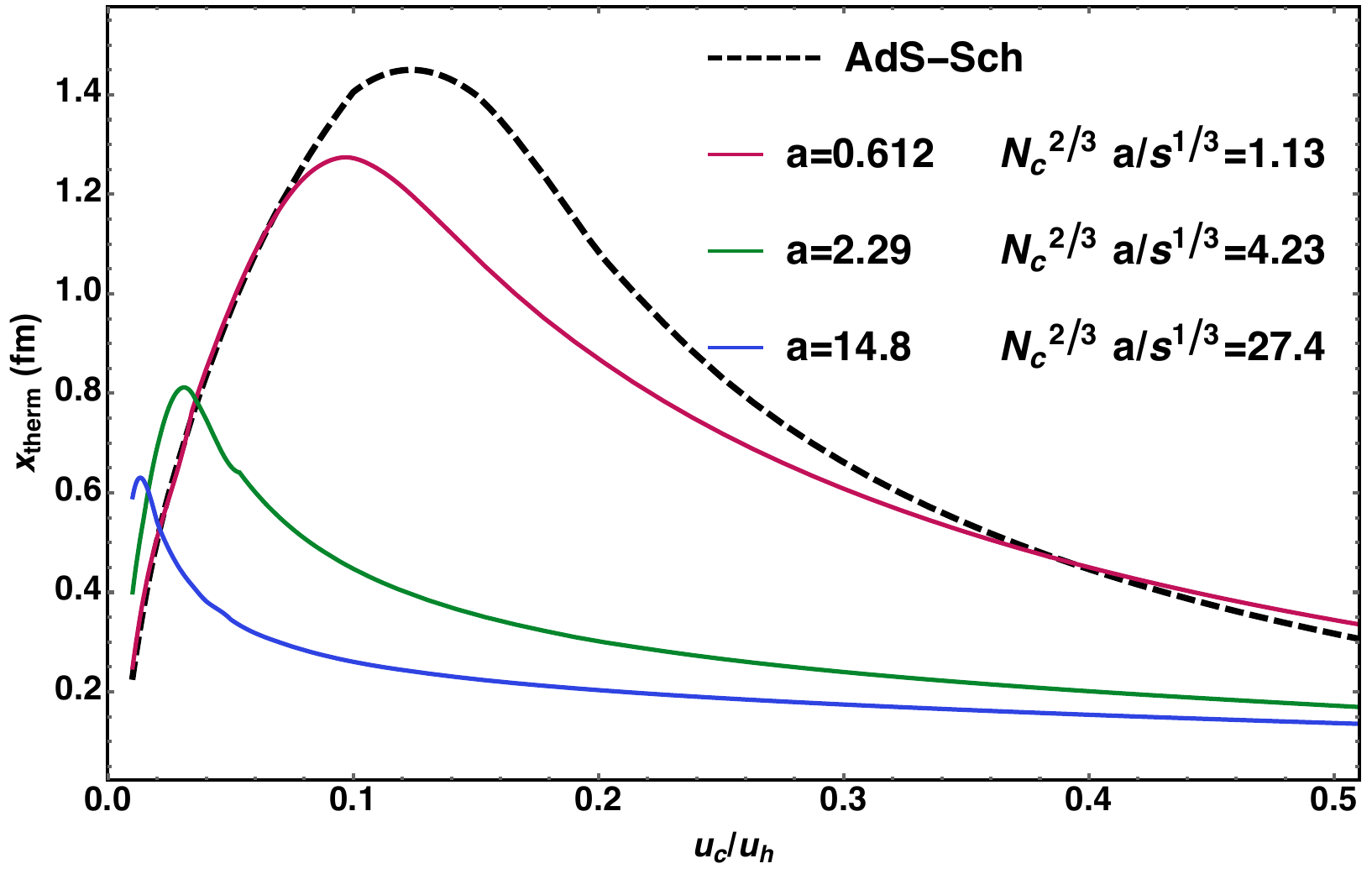}
\end{subfigure}
\begin{subfigure}[b]{.05in}
    \captionsetup{skip=-15pt,margin=-20pt}
    \caption{}
    \label{ConstantSb}
\end{subfigure}
\begin{subfigure}[b]{2.8in}
    \centering
    \includegraphics[width=2.8in]{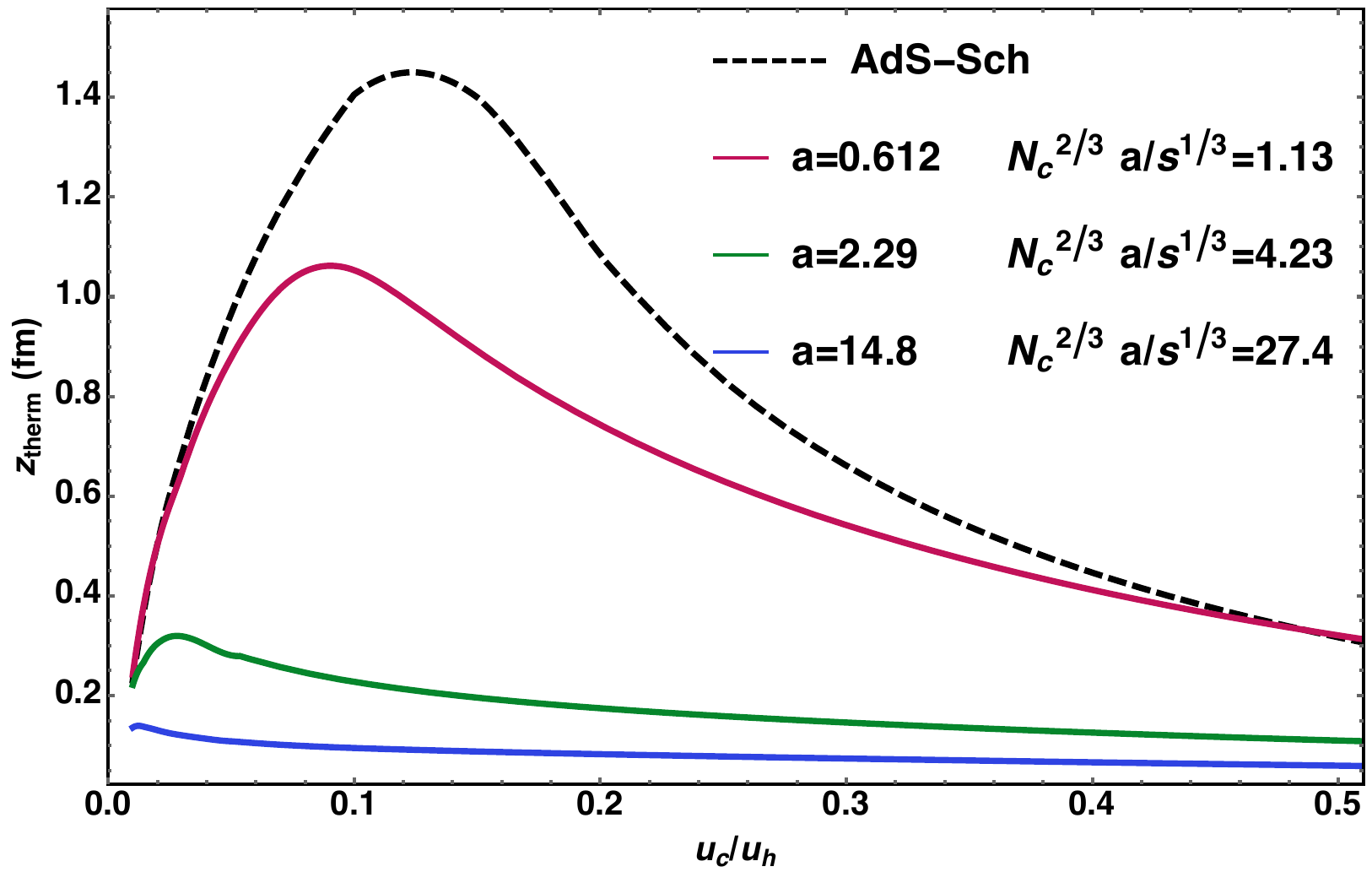}
\end{subfigure}
\vspace{.05in}
\caption{(Color online) The maximum stopping distance of the jet moving (\subref{ConstantSa}) in the x-direction and (\subref{ConstantSb}) in the z-direction in the anisotropic background with different anisotropy parameters. The anisotropic plasma has the same entropy density as the isotropic plasma, $N_c^{-2}~s= 0.159~GeV^3$.  In all cases, the quark has 100 GeV energy, but created at different radial distance from the boundary. The black dashed line is the stopping distance in the AdS-Sch background. }
\label{ConstantS}
\end{figure}

In \figtwo{ConstantT}{ConstantTa} , we plot the stopping distance of the light quark moving in x-direction in the anisotropic plasmas with different anisotropy parameter and compare it with the stopping distance of the quark with the same energy in the AdS-Sch background (Dashed-black line). In \figtwo{ConstantT}{ConstantTb} , the same quantity is plotted when the string is moving in the longitudinal direction in the anisotropic medium. One finds that increasing the anisotropy parameter leads to decreasing the stopping distance. This feature also appears when two plasmas have the same entropy density but different temperature, \fig{ConstantS}. In this figure, the stopping distance of the quark moving in both x-direction \figtwo{ConstantS}{ConstantSa} and in z-direction \figtwo{ConstantS}{ConstantSb} is plotted and compared with the case of isotropic plasma. Again, we suppose that in all cases the quark has $100$ GeV energy. The results are qualitatively similar, in both cases the jet quenching is more strong in the z-direction (beam direction) and also is more strong at the medium with larger anisotropy.

\subsection{Charged anisotropic background}%
\label{subsection:StoppingdistanceP}%


Now we study stopping distance of a light quark in a strongly coupled anisotropic plasma that carries a U(1) charge density. The drag force on a heavy quark in this background has been studied in \cite{Chakraborty:2014kfa,Cheng:2014fza}. It is found that the drag force in charged plasma is always larger than than the neutral plasma \cite{Fadafan:2008uv}. However in the charged anisotropic plasma, it depends on the value of $Q$ and $a$. For discussion on the jet quenching parameter in this background see \cite{Wang:2016noh}.

\begin{figure}[!htbp]
\centering
\begin{subfigure}[b]{.05in}
    \captionsetup{skip=-15pt,margin=-10pt}
    \caption{}
    \label{ChemicalPa}
\end{subfigure}
\begin{subfigure}[b]{2.8in}
    \centering
     \includegraphics[width=2.8in]{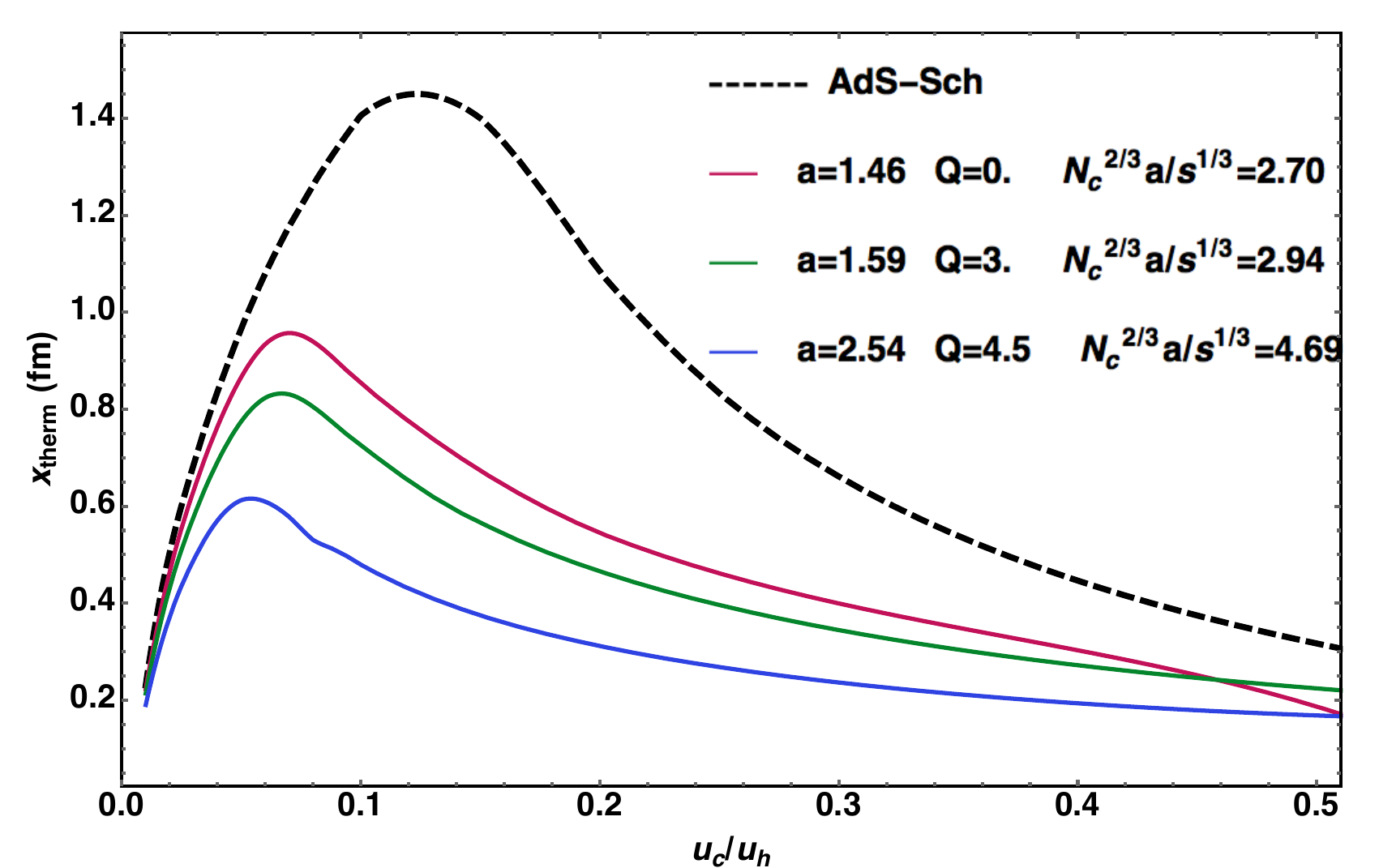}
\end{subfigure}
\begin{subfigure}[b]{.05in}
    \captionsetup{skip=-15pt,margin=-20pt}
    \caption{}
    \label{ChemicalPb}
\end{subfigure}
\begin{subfigure}[b]{2.8in}
    \centering
    \includegraphics[width=2.8in]{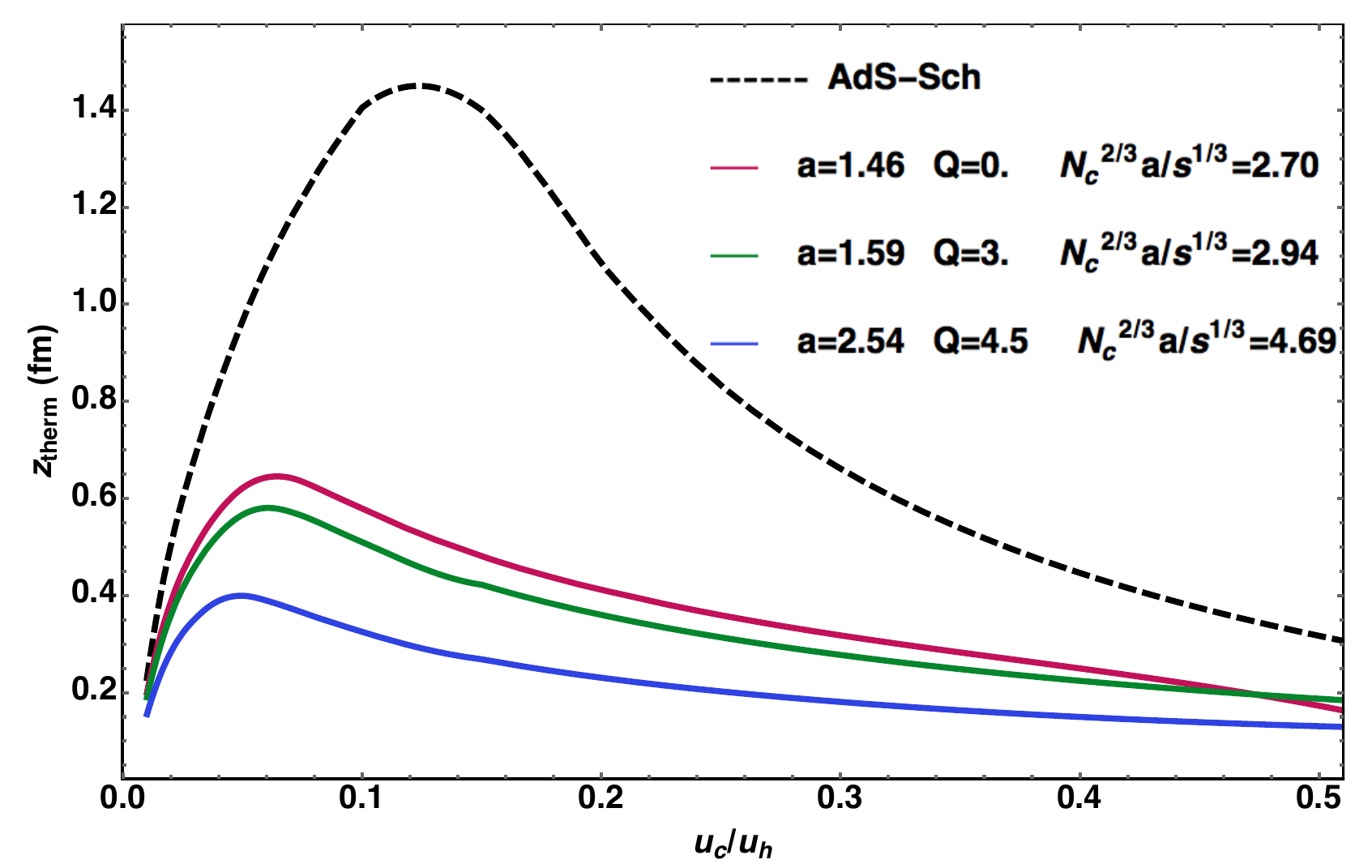}
\end{subfigure}
\vspace{.05in}
\caption{(Color online) The maximum stopping distance of the jet moving (\subref{ChemicalPa}) in the x-direction and (\subref{ChemicalPb}) in the z-direction the charged anisotropic background with different chemical potential. The value of $Q$ is 0, 3 and 4.5 for the purple, green and blue lines respectively. The charged anisotropic plasma has the same entropy density as the isotropic plasma, $N_c^{-2}~s= 0.159~GeV^3$. In all cases, the quark has 100 GeV energy, but created at different radial distance from the boundary.  }
\label{ChemicalP}
\end{figure}

The maximum stopping distance of quark in the charged anisotropic plasma \cite{Cheng:2014qia} as a function of $u_c/u_h$ is plotted in \fig{ChemicalP}. In this case, the quark has $100$ GeV energy and moves in x-direction (\figtwo{ChemicalP}{ChemicalPa}) and in z-direction (\figtwo{ChemicalP}{ChemicalPb}) of the charged anisotropic plasma with different chemical potentials. The results are compared with the case of isotropic plasma with the same entropy density. Again, the quantitative behavior of the jet is similar to the case of zero-chemical potential anisotropic plasma, although the quenching is more strong even for the intermediate anisotropy parameter (\fig{ChemicalP}).

\section{Conclusions}
\label{section:Conclusions}

In this paper, we have studied the effect of anisotropy on an energetic jet traveling through the QGP. We have considered a point-like initial condition string which moves through the anisotropic background. The equation of motion for the string have been solved numerically and plotted in \fig{string} for three typical strings moving in different directions in the $x-z$ plane. Our results show that the string highly suppressed in the longitudinal direction with respect to the transverse direction.

Although, the maximum stopping distance is not a good quantity to compute the observables like nuclear modification factor, but still is a good quantity to estimate the stopping power of the plasma. In \fig{fig2}, we have plotted the ratio of maximum stopping distance in anisotropic plasma to the same quantity in isotropic plasma for an energetic light quark with $P_q=0.99 E_q$ which moves in different directions in the plasma where the x-direction is plotted in red and the z-direction is plotted in purple. We have seen that the stopping distance decreases in both direction even at small anisotropy. The same quantity was evaluated using by a massless particle falling along the null geodesic in the WKB approximation in \cite{Muller:2012uu}. They also found the enhancement of jet suppression, although compare to our results coming from falling string their suppression is small. We have plotted the same quantity versus the direction of motion for the anisotropic medium with the same temperature or entropy density as isotropic one in \fig{Rx}. Our results demonstrate that the stopping distance always decreases in the presence of anisotropy, while the suppression of quark is less prominent once the two plasmas have the same entropy density.

We have found that the stopping distance is very sensitive to the IC of falling string, but still for any energy we can compute the maximum stopping distance. It was found that the stopping distance of the quark in the AdS-Sch metric is related to its energy as \eq{XE}. We have investigated this relation using our numerical results in the presence of anisotropy, \fig{X-E}. Our results show that the power of $1/3$ gets larger in the anisotropic medium. The same study has been done before for the non-conformal background and they also found this power larger than $1/3$ at small energies $\sim \,100\,GeV$ \cite{Ficnar:2011yj}.

We have plotted the stopping distance in terms of the string initial distance from the boundary for the strings with $100\, GeV$ energy. The qualitative behavior of the string in the both anisotropic (\fig{ConstantT}, and \fig{ConstantS}) and charged-anisotropic plasma (\fig{ChemicalP}) is the same as the isotropic plasma. However, the stopping distance decreases by increasing the anisotropy of the medium or the charge of the plasma.

Unlike the jet quenching parameter behavior in anisotropic plasma, we have found that the light quark energy loss and the stopping distance are always smaller than its counterpart isotropic plasma.

\textbf{Acknowledgments}
We thank M. Ali-Akbari and H. Soltanpanahi for useful discussions. KBF thanks U. Wiedemann and Y. Mehtar-Tani for the fruitful discussions on the anisotropic effects on the jet energy loss during at CERN. KBF acknowledges CERN TH Institute and University of Southampton for their hospitality. R. Morad wishes to thank the SA-CERN Collaboration, the South African National Research Foundation (NRF), and the National Institute for Theoretical Physics (NITheP) for their generous support.





\end{document}